\documentstyle[twocolumn,prl,aps,epsf]{revtex} 
\begin{document}
\title{On the Doping and Temperature Dependence of the Mass Enhancement 
Observed in the Cuprate Bi$_2$Sr$_2$CaCu$_2$O$_{8+\delta}$} 
\author{P. D. Johnson$^1$, T. Valla$^1$, A. V. Fedorov$^1$, Z. Yusof$^2$, B. O. 
Wells$^2$,  Q. Li$^3$, A. R. Moodenbaugh$^3$, G. D. Gu$^4$, N. Koshizuka$^5$, C. 
Kendziora$^6$, Sha Jian$^7$ and D. G. Hinks$^7$}
\address{$^1$ Department of Physics, Brookhaven National Laboratory, Upton, NY, 
11973-5000}
\address{$^2$ Department of Physics, University of Connecticut, 2152 Hillside 
Road U-46, Storrs, CT 06269}
\address{$^3$ Division of Materials Sciences, Brookhaven National Laboratory, 
Upton, NY 11973}
\address{$^4$ School of Physics, University of New South Wales, P.O. Box 1, 
Kensington, New South Wales, Australia 2033}
\address{$^5$ Superconductivity Research laboratory, ISTEC, 10-13, Shinonome I-
chrome, Koto-ku, Tokyo 135, Japan }
\address{$^6$ Materials Sciences Division, Naval Research Laboratory, Washington 
DC 20375}
\address{$^7$ Materials Sciences Division, Argonne National Laboratory, Argonne 
IL 60439}

\date{\today}
\address{ {\em \bigskip \begin{quote}
High-resolution photoemission is used to study the electronic structure of the 
cuprate superconductor, Bi$_2$Sr$_2$CaCu$_2$O$_{8+\delta}$, as a function of 
hole doping and temperature.  A kink observed in the band dispersion in the 
nodal or $(0,0)$ to $(\pi,\pi)$ direction in the superconducting state is 
associated with coupling to a resonant mode observed in neutron scattering.  
From the measured real part of the self energy it is possible to extract a 
coupling constant which is largest in the underdoped regime, then decreasing 
continuously into the overdoped regime.
\end{quote}}}
\maketitle
In any system, electrons (or holes) may interact strongly with various 
excitations resulting in modifications to both their lifetime and binding 
energy.  The quantity that describes these effects is the self-energy, 
$\Sigma({\bf k},\omega)$ the imaginary part representing the scattering rate or 
inverse lifetime, the real part, the shift in energy. An electron in a solid 
thus becomes "dressed" with a "cloud" of excitations, acquiring a different 
effective mass, but still behaving as a "single-particle" excitation or 
quasiparticle. This represents the traditional Fermi liquid (FL) picture. In 
more exotic materials, an electron or hole may lose its "single-particle" 
integrity and decay completely into collective excitations.

It has recently been demonstrated that angle-resolved photoemission (ARPES) is 
an excellent tool for momentum resolving self-energy effects.  A measurement of 
the self-energy allows the determination of the coupling strength as well as the 
identification of the energy scale of the fluctuations involved in the coupling.  
As an example, self-energy effects due to the electron-phonon interaction have 
been identified in molybdenum \cite{moly} and beryllium \cite{beril}. In these 
cases the 
energy scale is represented by the Debye energy.  More recently, in studies of 
the layered dichalcogenide 2H-TaSe$_2$, self-energy corrections reflecting 
coupling to fluctuations in the charge density wave order parameter have been 
observed \cite{tase}. The extension of such studies to the realm of high-Tc 
superconductivity allows the identification of the appropriate energy scales 
describing the fluctuations in these materials.  It is not clear however, that a 
Fermi liquid methodology is appropriate for the high temperature superconductors 
(HTSC).  Indeed in a recent study of optimally doped 
Bi$_2$Sr$_2$CaCu$_2$O$_{8+\delta}$, the non-Fermi liquid like nature of the 
material was demonstrated \cite{science}. It was shown that the imaginary part 
of the self-energy follows a Marginal Fermi liquid (MFL) behavior \cite{varma} 
with quantum critical scaling, suggesting the absence of any energy scale 
associated with the nodal excitations. However, in the same study 
\cite{science}, a change in the mass enhancement was observed in the 
superconducting state, indicating structure in the self-energy and the 
appearance of an energy scale well removed from the Fermi level. The 
corresponding change in the Im$\Sigma$ was not observed directly. Subsequent 
experimental studies have reported that in the superconducting state the mass 
enhancement exists over a large portion of the Fermi surface 
\cite{bogdanov,kaminski}.  Theoretical studies have focussed on the possibility 
that these observations reflect coupling to the magnetic resonance peak observed 
in neutron scattering studies \cite{eschrig-n}.

In this paper, we examine the doping and temperature dependence of the mass 
enhancement. We find that in the normal state the self-energy is well described 
within the framework of the MFL model. However upon entering the superconducting 
state, changes occur in the ARPES spectra. We find that the self-energy 
correction and associated mass enhancement are strongly dependent on the hole 
doping level, decreasing continuously with doping. Further the energy scale 
observed in the superconducting state is linearly dependent on the transition 
temperature $T_C$. This dependence is almost identical to that observed for the 
resonant collective mode observed in neutron scattering studies 
\cite{dai,keimer}.

The experimental studies reported in this paper were carried out on a 
photoemission facility equipped with a Scienta electron spectrometer \cite{PDJ}. 
In this instrument, the total spectral response may be measured as a function of 
angle and energy simultaneously with a momentum resolution of $\sim0.005$ 
\AA$^{-1}$ and an energy resolution of $\sim10$ meV. Photons were provided by a 
Normal Incidence Monochromator based at the National Synchrotron Light Source. 
Samples of optimally doped ($T_C = 91$ K) Bi$_2$Sr$_2$CaCu$_2$O$_{8+\delta}$ 
were produced by the floating zone method \cite{gu}. The underdoping and 
overdoping was achieved by annealing optimally doped samples in argon 
\cite{under} and in oxygen \cite{over}, respectively. All samples were mounted 
on a liquid He cryostat and cleaved {\it in-situ} in the UHV chamber with base 
pressure $3\times 10^{-9}$ Pa. The sample temperature was measured using a 
calibrated silicon sensor. The self-energy corrections were determined either 
from energy distribution curves (EDC) or from momentum distribution curves 
(MDC). The EDC represents a measure of the intensity as a function of binding 
energy at constant momentum and the MDC a measure of the intensity as a function 
of momentum at constant binding energy. 

In Fig. 1 we show the photoemission intensities recorded as a function of 
binding energy and momentum in the (0,0) to ($\pi,\pi$) direction of the 
Brillouin zone for from left to right, the underdoped (UD), optimally doped (OP) 
and overdoped (OD) Bi$_2$Sr$_2$CaCu$_2$O$_{8+\delta}$ samples, all in the 
superconducting state. In the lower panel we show the corresponding band 
dispersions obtained from MDC's for the superconducting and normal states.  It 
is clear that even in the normal state, the dispersion in the vicinity of the 
Fermi level deviates from the linear dispersion predicted by first-principles 
band structure calculations \cite{15a}. In the superconducting state, an 
additional modification to the dispersion develops for the under- and optimally 
doped samples. In the overdoped material there is no detectable change in 
dispersion.

The spectra in Fig. 1 display other interesting features. For $\omega\ge50$ 
meV the velocity, or rate of band dispersion decreases with doping. This is 
surprising because one might anticipate the velocity decreasing as the states 
became more localized in the underdoped regime. Secondly, the spectral response 
is less well defined in the underdoped regime with the spectral width in both 
energy and momentum exceeding the amount of dispersion from $k=k_{F}$.   
Following recent suggestions of Orgad {\it et al}. \cite{orgad} these two 
observations may be evidence of increased electron (hole) fractionalization in 
the underdoped regime.

In Fig. 2 we show the deviation, Re$\Sigma$, from the non-interacting 
dispersion as a function of binding energy for the three doping levels for both 
the superconducting and normal states.  We define the non-interacting dispersion 
as a straight line that coincides with the experimentally measured Fermi wave-
vector and the dispersion measured at a higher binding energy, typically 250 
meV, as indicated in Fig. 1. The real part of the self-energy is set to zero 
at these points. The overall magnitude of the self-energy continuously decreases 
with doping in both the normal and superconducting state. We also show the 
difference in Re$\Sigma$ between the superconducting state and normal state. We 
suggest that this difference represents a change in the excitation spectrum 
associated with the coupling upon entering the superconducting state. This 
change should also be manifested in measurements of Im$\Sigma$. While apparent 
in the underdoped regime \cite{PRB}, the effect is too subtle to 
observe directly in the optimally doped material.

In Fig. 3 we plot different quantities deduced from Re$\Sigma$ as a function
of the deviation from the maximum $T_C=91$ K, characterizing the optimally doped material.
In panel (a) we plot both $\omega_0$, the energy of 
the maximum in Re$\Sigma$ in the superconducting state and $\omega_{0}^{sc}$, 
the energy of the maximum in the difference between the superconducting- and 
normal-state. Since in the overdoped regime the difference between the 
superconducting- and normal- state dispersion vanishes, the energy scale characterizing
the superconducting state could not be detected in the nodal region. The characteristic energy 
was therefore identified only for a limited range of overdoping and 
only by moving away from the node towards the ($\pi$,0) region where the coupling is observed to
be stronger while the characteristic energy of the kink remains momentum-independent \cite{kaminski,FS}. 
Indeed, measurements of the renormalized velocity in the superconducting state indicate that, for optimal doping, 
the coupling increases 
by a factor of 3 or more on moving towards the ($\pi$,0) region \cite{FS}.
In the overdoped regime, there is also more uncertainty in the transition temperature
due to the increased tendency of losing oxygen \cite{exp-od}. 

In the underdoped regime it is clear that the characteristic energies $\omega_{0}$ and 
$\omega_{0}^{sc}$ scale linearly with $T_C$ as opposed to, for instance the 
magnitude of the maximal gap observed in these materials. The latter increases 
continuously on going into the underdoped regime \cite{campu}. When fitted with 
a straight line, the data points for $\omega_{0}^{sc}$ and $\omega_{0}$ in 
Fig. 3(a) yield $\omega\sim6k_{B}T_{C}$. This behavior is reminiscent of that 
reported in neutron scattering studies where all the characteristic low-energy features in
the superconducting state appear to scale with $T_C$. In particular, the 
resonance energy, $E_{r}$, was found to scale as $E_{r}\sim5.4k_{B}T_{C}$ \cite{keimer}, while
the spin gap scales as $\Delta_{s}\sim3.8k_{B}T_{C}$ \cite{spin-gap}.  The possibility that the kink
reflects coupling to zone boundary longitudinal optical phonons has also been discussed in the literature 
\cite{shen}. However, recent neutron studies indicate that these phonons ocurr at the same energy, independent of doping \cite{phonons}.
This is in complete contrast to the doping dependence of the kink energy observed in the present study.

The real part of the self energy may also be used to extract the coupling 
strength to the excitations involved in the coupling \cite{mahan} via 
$\lambda=-(\partial{\mathrm Re}\Sigma/\partial\omega)_{E_F}$, neglecting momentum dependence of the self 
energy in the narrow interval around $k_F$. The coupling constant $\lambda$ is 
simply obtained by fitting the low energy part of Re$\Sigma(\omega)$ to a 
straight line. In Fig. 3(b) we plot $\lambda$ for different samples in the 
superconducting state as a function of $T_{C}$. From the figure we see that the 
coupling decreases continuously with increasing doping level, reflecting the 
latter rather than the transition temperature.

Shown in Fig. 4, the "kink" and the magnetic resonance mode also display the 
same temperature dependence. Here the temperature dependence of Re$\Sigma$ for 
the underdoped (UD69K) sample, measured at the characteristic energy 
$\omega=\omega_{0}^{sc}$, is compared with the intensity of the resonance mode 
measured in inelastic neutron scattering (INS) from an YBa$_2$Cu$_3$O$_{6+x}$ 
sample with similar $T_C$ \cite{dai}. Although our comparison is between two different systems, 
we note that a recent INS study combined with ARPES on the same underdoped Bi$_2$Sr$_2$CaCu$_2$O$_{8+\delta}$ 
sample ($T_C=70$ K) have reported results nearly identical to the present study \cite{mesot}. The identical temperature dependence in Fig. 4 
points to a common origin for both phenomena. Note that both features exist at temperatures significantly higher 
than $T_C$. The temperature range over which 
they lose intensity at the fastest rate appears close to $T_C$. However the 
features show intensity up to temperatures close to $T^*$, the pseudo-gap 
temperature measured in various transport properties. 

We have provided strong evidence that in the superconducting state the low-energy excitations are 
affected by the low-energy part of spin fluctuation spectrum observed in neutron scattering. The question naturally arises as to what is 
responsible for the mass enhancement observed in the normal state for all samples. If phonons 
were the source of coupling, we might anticipate a {\it saturation in the 
scattering 
rate at frequencies greater than the Debye frequency and a marked temperature 
dependence in that range} \cite{moly,beril,mahan}. This is clearly in contrast 
with optical 
conductivity \cite{puchkov} and photoemission \cite{science,bogdanov,kaminski} 
experiments on the 
optimally- and under- doped samples, where the obvious lack of saturation in the 
scattering rate points to the absence of a well-defined cutoff in the excitation 
spectrum. Phonons, on the other hand, are {\it always} limited to a finite 
energy range (usually $\le 100$ meV in these materials). Further, as we have noted earlier, 
no significant temperature and doping dependence is observed in the phonon spectrum. More 
likely is the possibility that the enhancement still reflects coupling to spin 
excitations and that the spectrum of excitations changes with temperature. 
Indeed, the spin response changes between the normal and (pseudo)gapped 
state. A spin gap opens at low energies in superconducting state, and the strong resonance mode appears at 
commensurate momenta, ${\bf Q}=(\pi,\pi)$. The spectrum becomes incommensurate for lower and higher 
energies. These temperature changes in the susceptibility are in one-by-one 
correlation with the temperature changes in the single-particle spectrum. The 
stronger the temperature change in susceptibility, the stronger the temperature 
change in photoemission. This suggests that in the overdoped regime, not only 
the susceptibility weakens, but also that the change between the normal- and
superconducting- states gradually disappears. In systems where the "resonance mode" dominates the 
susceptibility in the superconducting state, coupling to it causes the "kink" in the 
dispersion at roughly the same energy \cite{einstein}. If in the normal state
the coupling still reflects spin fluctuations, it should scale with the spin 
susceptibility $\chi^{\prime\prime}$. Indeed, the spin susceptibility has been 
shown to increase in the underdoped region in both the normal and 
superconducting states \cite{bourges}, consistent with the present findings. As 
a two-particle response function, it is only limited by the band width, in 
accord with the absence of a clear cutoff in the single-particle scattering rate. 

The Kramers Kronig transform of a scattering rate, Im$\Sigma\propto\omega$ gives  
Re$\Sigma=g'\omega\ln(\omega_{c}/\omega)$, where $g'$ is a coupling constant and 
$\omega_{c}$ is a high-energy cutoff as defined within the MFL framework 
\cite{varma}. Indeed, any system with the scattering rate linear in binding 
energy should also display a logaritmic correction to the dispersion. Fitting 
the Re$\Sigma$ shown in the Fig. 2 gives a value of 
approximately 230 meV for $\omega_{c}$ and values for $g'$ of 0.54, 0.29 and 
0.24 for the underdoped, optimally doped and overdoped samples respectively. We 
note that the value obtained for $g'$ for the optimally doped material is 
consistent with the $\omega$ dependence of Im$\Sigma$ found in our earlier study 
\cite{science} and is in good agreement with values obtained in a recent 
analysis of normal state EDCs by Abrahams and Varma \cite{PNAS}. The observation 
that the MFL form for the self energy fits the data in the overdoped region in 
both the normal and superconducting state is suggestive that the quantum 
critical point, if it exists, may well be displaced towards the overdoped 
region. 

To conclude, the doping, temperature and momentum dependences of the various 
characteristic features seen in the cuprate photoemission spectra are consistent 
with a picture where the injected hole couples to the spin fluctuations observed 
in INS. Indeed this evidence showing that the mass enhancement associated with 
the opening of the (pseudo)gap is electronic in nature is reminiscent of the 
behavior reported for 2H-TaSe$_2$ \cite{tase}. In the latter study a mass-
enhancement or kink in the dispersion developed with the formation of the CDW 
gap. The energy scale, being much larger than the Debye energy, was attributed 
to electronic excitations across the CDW gap. In the present study, from the 
observation that in the underdoped region the transition temperature $T_C$ 
decreases as the coupling increases, we conclude that the coupling strength 
alone is clearly insufficient to explain the superconductivity in these 
materials and that some other ingredients, such as the carrier concentration 
\cite{PRB,ding} and 
the phase coherence \cite{kivelson,forro}, are clearly required.  

The authors would like to acknowledge useful discussions with Alexei Tsvelik, 
Andy Millis, V.N. Muthukumar, Dimitri Basov, Andrey Chubukov, John Tranquada, 
Steve Kivelson and Takeshi Egami. The work was supported in part by the Department of Energy under 
contract number DE-AC02-98CH10886 and in part by the New Energy and Industrial 
Technology Development Organization.

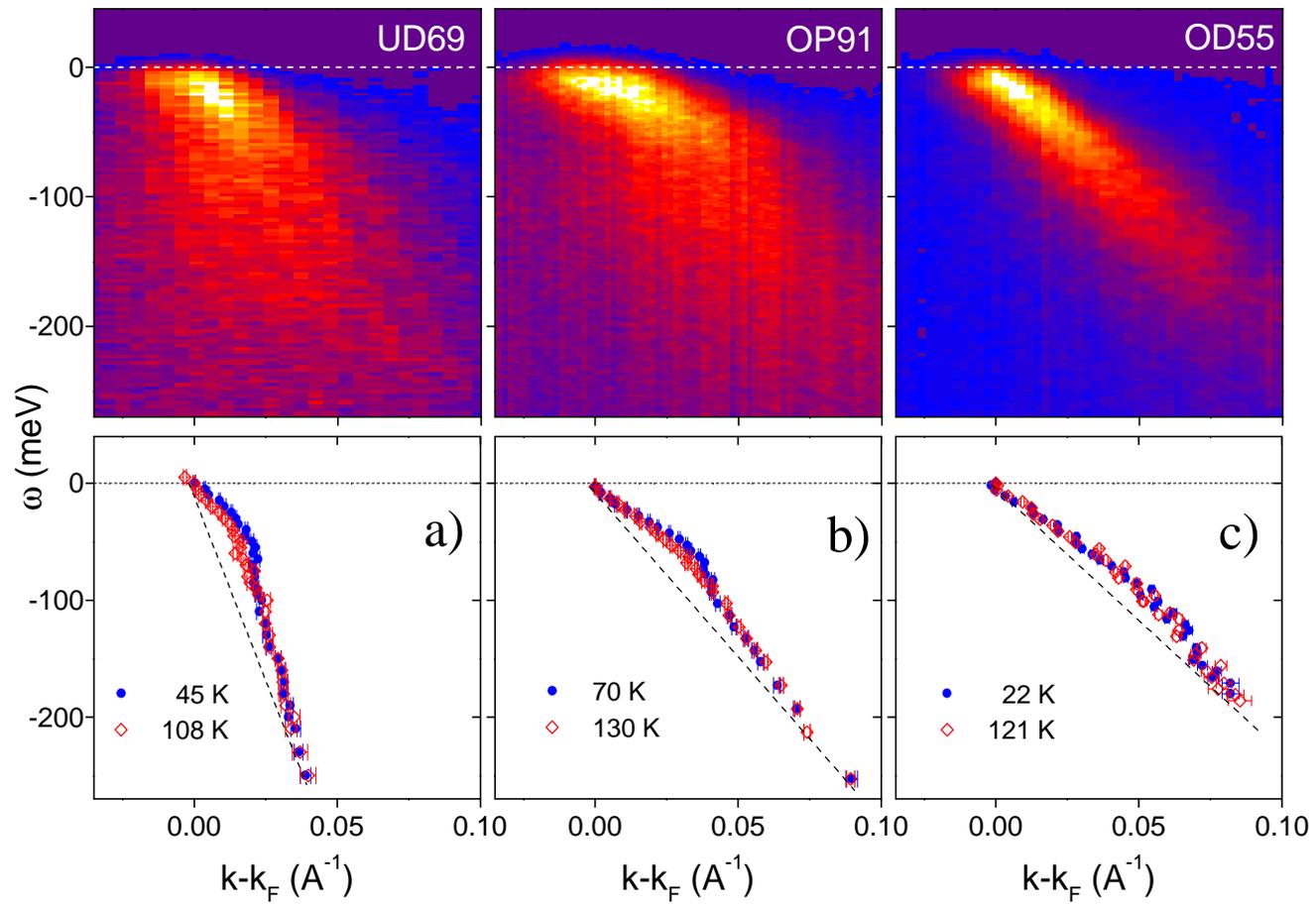
\begin{figure}
\caption{
Upper panels:- Two dimensional photoemission intensities observed from (a) 
underdoped (UD), (b) optimally doped (OP) and (c) overdoped (OD) 
Bi$_2$Sr$_2$CaCu$_2$O$_{8+\delta}$ samples. The superconducting transition 
temperatures are indicated.  Lower panels:- The dotted lines indicate the MDC 
deduced dispersions for both the superconducting (blue dots) and normal states 
(open red diamonds) corresponding to the different samples in the panels above.}
\label{fig:1}
\end{figure}

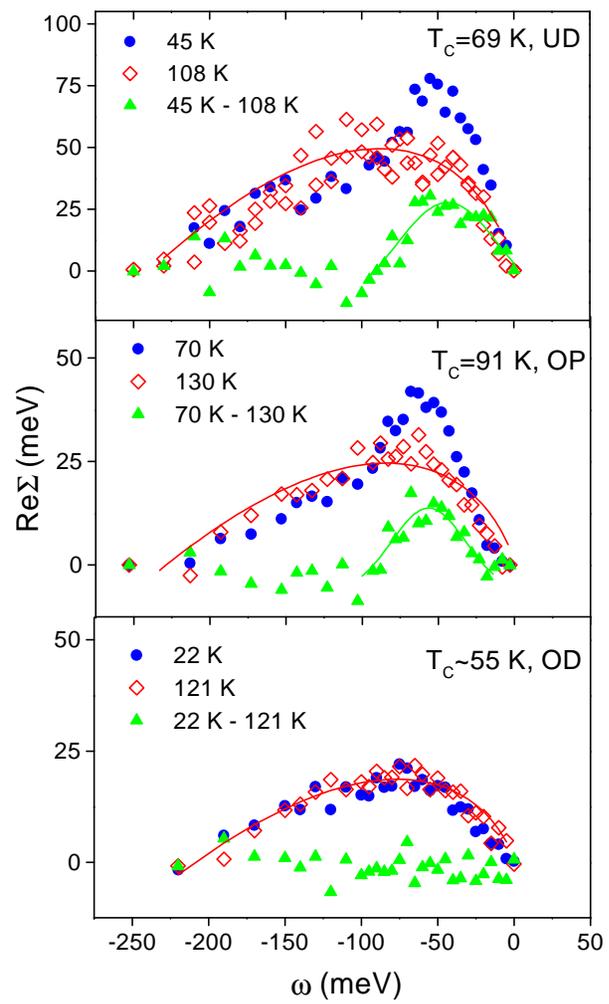
\begin{figure}
\caption{
Re$\Sigma$ as a function of binding energy for the superconducting (blue dots) 
and normal states (open red diamonds) for the underdoped (UD), optimally doped 
(OP) and overdoped (OD) samples, as indicated. The solid lines through the 
normal state data represent MFL fits to the data.  The difference between the 
superconducting and normal Re$\Sigma$ for each level of doping is also plotted 
(green triangles).  The lines through the latter are    Gaussian fits to extract 
the peak energy $\omega_{0}^{sc}$.}
\label{fig:2}
\end{figure}

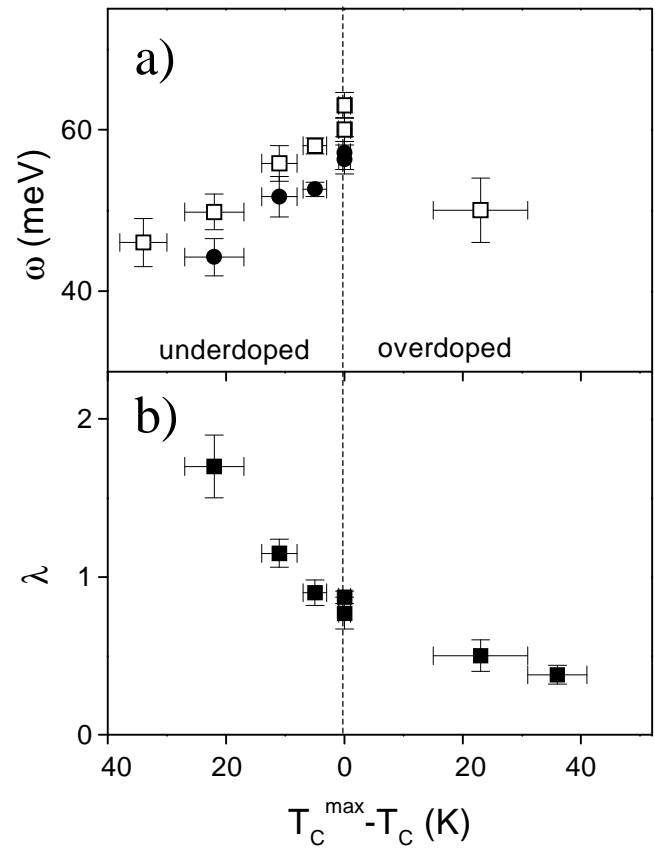
\begin{figure}
\caption{
(a)  Plot of $\omega_{0}$, the energy of the maximum value of Re$\Sigma$ in the 
superconducting state (open squares), and $\omega_{0}^{sc}$ (solid circles), the 
energy of the maximum in difference between the superconducting and normal state 
values plotted as a function of $T_C$ referenced to the maximum $T_C^{max}$ 
($\sim91$ K).
(b)  The coupling constant $\lambda$, determined as described in the text and 
plotted as a function of $T_C$.}
\label{fig:3}
\end{figure}

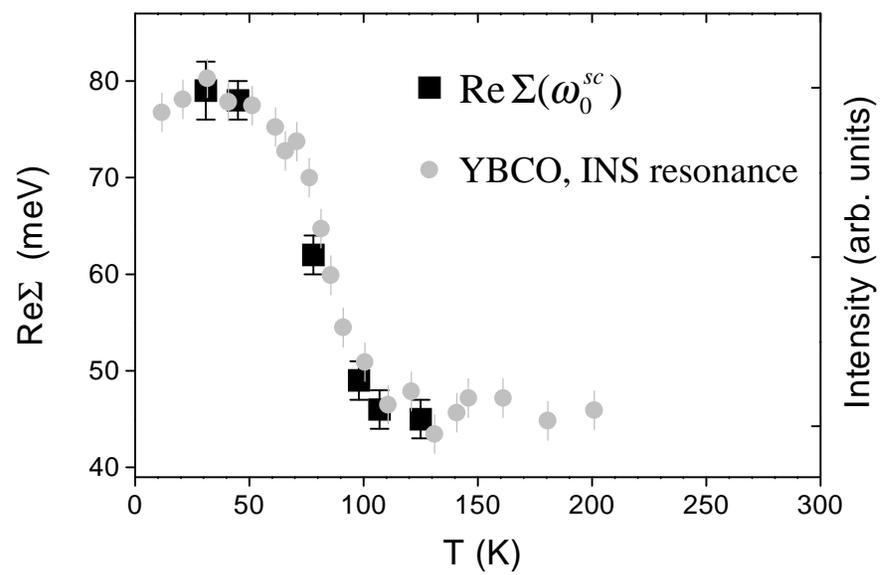
\begin{figure}
\caption{
Temperature dependence of Re$\Sigma(\omega_0^{sc})$ from the nodal line for the 
UD69K sample (black squares) compared with the temperature dependence of the 
intensity of the commensurate resonance mode observed in neutron scattering 
studies of underdoped YBa$_2$Cu$_3$O$_{6+x}$, $T_C=74$ K, (ref. [9]) (gray 
circles).}
\label{fig:4}
\end{figure}


\begin{references}

\bibitem{moly} T. Valla, A. V. Fedorov, P. D. Johnson and S. L. Hulbert, Phys. 
Rev. 
Lett. {\bf 83}, 2085 (1999).

\bibitem{beril} M. Hengsberger {\it et al}, Phys. Rev. Lett. {\bf 
83}, 592 (1999); S. LaShell, E. Jensen and T. Balasubramanian, Phys. Rev. B {\bf 
61}, 2371 (2000).

\bibitem{tase} T. Valla {\it at al}, Phys. Rev. Lett. {\bf 85}, 4759 (2000).

\bibitem{science} T. Valla {\it at al}, Science {\bf 285}, 2110 (1999).

\bibitem{varma} C. M. Varma {\it at al}, Phys. Rev. Lett. {\bf 63}, 1936 (1989).

\bibitem{bogdanov} P. V. Bogdanov {\it at al}, Phys. Rev. Lett. {\bf 85}, 2581 
(2000).

\bibitem{kaminski}  A. Kaminski {\it at al}, Phys. Rev. Lett. {\bf 86}, 1070 
(2001).

\bibitem{eschrig-n} M. Eschrig and M. R. Norman, Phys. Rev. Lett. {\bf 85}, 3261 
(2000).

\bibitem{dai} P. Dai {\it et al}, Science {\bf 284}, 1346 (1999).

\bibitem{keimer} H. F. Fong {\it et al}, Nature {\bf 398}, 588 (1999); H. He 
{\it et al}, Phys. Rev. Lett. {\bf 86}, 1610 (2001).

\bibitem{PDJ} P. D. Johnson {\it et al}, Proceedings of the $11^{th}$ Synch. 
Rad. Conf. AIP Press NY, {\bf 521}, 73 (2000).

\bibitem{gu} G. D. Gu, K. Takamaku, N. Koshizuka and S. Tanaka, J. Crystallogr. 
Growth {\bf 130}, 325 (1990); N. Miyakama {\it et al}, Phys. Rev. Lett. {\bf 
80}, 157 (1998).

\bibitem{under} A. R. Moodenbaugh, D. A. Fisher, Y. L. Wang and Y. Fukumoto, 
Physica C {\bf 268}, 107 (1996).

\bibitem{over} C. Kendziora, R. J. Kelley, E. Skelton, and M. Onellion, Physica 
C {\bf 257}, 74 (1996). 

\bibitem{15a} H. Krakauer and W. E. Pickett, Phys. Rev. Lett. {\bf 60}, 1665 
(1988).

\bibitem{orgad} D. Orgad {\it et al}, cond-mat/0005457.

\bibitem{PRB} T. Valla {\it et al}, to be published.

\bibitem{FS} T. Valla {\it et al}, Phys. Rev. Lett. {\bf 85}, 828 (2000).

\bibitem{exp-od} To minimize uncertainties for the overdoped samples, $T_C$ was 
measured in situ, by monitoring the superconducting gap and by measuring the 
susceptibility before and after the photoemission.

\bibitem{campu} J. C. Campuzano {\it et al}, Phys. Rev. Lett. {\bf 83}, 3709 
(1999).

\bibitem{spin-gap} P. Dai, H. A. Mook, R. D. Hunt and F. Do{\u g}an, Phys. Rev. B {\bf 63}, 054525 (2001).

\bibitem{shen} A. Lanzara {\it et al}, cond-mat/0102227; Z.-X. Shen, A. Lanzara and N. Nagaosa, cond-mat/0102244.

\bibitem{phonons} Y. Petrov {\it et al}, cond-mat/0003414; R. J. McQueeney {\it et al}, cond-mat/0105593.

\bibitem{mahan} G. D. Mahan, {\it Many Particle Physics} (Plenum Press, New York 
1991).

\bibitem{mesot} J. Mesot {\it et al}, cond-mat/0102339.

\bibitem{puchkov} A. V. Puchkov, D. N. Basov, and T. Timusk, J. Phys. Condens. 
Matter {\bf 8}, 10049 (1996).

\bibitem{einstein} Coupling to an Einstein-like mode results in a peak in the 
real part of the self energy at that energy.

\bibitem{bourges} P. Bourges, in {\it The Gap symmetry and Fluctuations in High 
Temperature Superconductors}, ed. By J. Bok, G. Deutscher, D. Pavuna, and S.A. 
Wolf (Plenum Press, 1998).

\bibitem{PNAS} E. Abrahams and C. Varma, Proc. Natl. Acad. of Sciences {\bf 97}, 
5714 (2000).

\bibitem{ding} H. Ding {\it et al}, cond-mat/0006143.

\bibitem{kivelson} V. J. Emery and S. A. Kivelson, Nature {\bf 374}, 434 (1995).

\bibitem{forro} I. Vobornik {\it et al}, Phys. Rev. Lett. {\bf 82}, 3128 (1999); 
Phys. Rev. B {\bf 61}, 11248 (2000).

\end{references}
\end{document}